\documentclass{aa}
\usepackage{graphicx}
\usepackage{natbib}
\usepackage{txfonts}

\def\LS{LS~5039}
\def\PSRB{PSR~B1259-63/LS~2883}

\def\PSRJ{PSR~J2032+4127/MT91 213}
\def\HESSJ{HESS~J0632+057}
\def\LSI{LS~I+61$^{\circ}$303}

\begin{document}

\title{Radio absorption in high-mass gamma-ray binaries}

\author{A. M. Chen\inst{1} \and Y. D. Guo\inst{2} \and Y. W. Yu\inst{2} \and J. Takata\inst{1}}
\institute{School of Physics, Huazhong University of Science and Technology, Wuhan 430074, China;
\and Institute of Astrophysics, Central China Normal University, Wuhan 430079, China}
\mail{chensm@mails.ccnu.edu.cn; yuyw@mail.ccnu.edu.cn; takata@hust.edu.cn }

\date{Received xxx / Accepted xxx}

\abstract{
High-mass gamma-ray binaries consist of a presumptive pulsar in orbit with a massive star.
The intense outflows from the star can absorb radio emission from the pulsar, making the detection of pulsation difficult. In this work, we present the basic geometry and formulae that describe the absorption process of a pulsar in binary with an O/B star and apply our model to two typical and well-studied binaries: PSR~B1259-63/LS~2883 and LS 5039.
We investigate the influences of the equatorial disc of LS 2883 with different orientations on the dispersion measure and free-free absorption of the radio pulsation from PSR B1259-63. The observed data are consistent with the disc inserted on the orbital plane with a relatively large inclination angle.
For LS 5039, due to its tight orbit, it was believed that the strong wind absorption makes detecting radio emissions from the putative pulsar unlikely. However, considering the wind interaction and orbital motion, a bow shock cavity and a Coriolis shock would be formed, thereby allowing the pulsations to partially avoid stellar outflow absorption. We investigate the dependence of the radio optical depth on the observing frequencies, the orbital inclination angle, and the wind parameters. We suppose that the presumptive pulsar in LS 5039 is similar to PSR B1259-63 with pulsed emission extending to several tens of gigahertz. In that case, there could be a transparent window for radio pulsations when the pulsar is moving around the inferior conjunction.
The following deep monitoring of LS 5039 and other systems by radio telescopes at high radio frequencies might reveal the nature of compact objects in the future. Alternatively, even a null detection could still provide further constraints on the properties of the putative pulsar and stellar outflows.

}

\authorrunning{Chen et al.}
\titlerunning{Radio absorption in HMGBs}

\maketitle

\section{Introduction}
High-mass gamma-ray binaries (HMGBs), consisting of a compact object and a massive main-sequence star, emit broad-band radiations from radio to $\gamma$-rays (see Dubus 2013;
Paredes \& Bordas 2019 for reviews on HMGBs). The energy spectra of HMGBs peak above 1 MeV, distinguishing them from well-known high-mass X-ray binaries (HMXBs). So far, the proposed emission models for HMGBs can be divided into the following two classes: (1) in the micro-quasar model, the broad-band emission of HMGBs arises from the relativistic jet produced by the accretion process of the compact object (e.g. Bosch-Ramon \& Khangulyan 2009); and (2) in the pulsar model, the emission is attributed to the termination shock that formed by wind interactions between the pulsar and the star (e.g. Dubus 2006).
In comparison, the latter scenario seems to be favoured by growing evidence, such as the spectral and timing characteristics as well as the lack of accretion signals in HMGBs.
Detecting pulsation signals from these systems would undoubtedly bring a conclusion to the arguments on the nature of compact objects in these HMGBs. Furthermore, timing observations can help to constrain the orbital parameters and the properties of the presumptive pulsar.

Until now, only nine HMGBs have been detected, and only two of them with compact objects are identified as radio pulsars, namely, \PSRB\ and \PSRJ\ (Johnston et al. 1992; Lyne et al. 2015). Recently, Yoneda et al. (2020) have reported the detection of X-ray pulsations in LS 5039 with a period of $9$ s. However, the timing analysis of Volkov et al. (2021) shows that the statistical significance of the pulsation might be too low to attribute it to the pulsar spin period.
Several tentative observations have been performed to search for radio pulsations in LS 5039 and other HMGBs, but each concluded with a null detection (e.g. McSwain et al. 2011; Ca{\~nellas et al. 2012). As a natural explanation, the result of a non-detection could be just because the pulsar emission beam is not pointing towards us.
Additionally, the intense outflows from the massive companion could further obscure the radio pulsations through free-free absorption (FFA), even if the line of sight (LOS) is in the pulsar radio beam. To minimise the influence of this wind obscuration, previous radio observations of HMGBs were usually set around the apastron of their orbits, where the putative pulsar encounters the most negligible wind density. However, due to the tight orbits of these HMGBs, the stellar wind can always be optically thick, even at the apastron. Furthermore, for some Be/$\gamma$-ray binaries, the presence of a stellar disc results in much more complicated absorptions of radio pulsations.

Radio absorption processes in HMGBs have been studied in previous works, where the effects of pulsar wind cavities are usually ignored (e.g. McSwain et al. 2011; Ca{\~n}ellas et al. 2012; Koralewska et al. 2018).
Here, we take a more proper binary geometry into account and show that minimum absorption could occur around the inferior conjunction (INFC). As the pulsar wind collides with companion outflows, a bow shock cavity is formed, shielding the pulsar from dense stellar wind (e.g. Dubus 2006). In addition, a Coriolis shock is created in the opposite direction from the star due to the fast orbital motion of the pulsar (Bosch-Ramon \& Barkov 2011).
When the pulsar moves around the INFC, the LOS could point to the pulsar through the Coriolis shock and pulsar wind cavity if the binary orbit is appropriately inclined. Depending on the stellar outflow properties, radio emission from the putative pulsar could become detectable around the INFC under certain conditions. For a specific HMGB such as LS 5039, a detailed investigation of its radio absorption process would be beneficial for searching for possible radio pulsations in the future and identifying the nature of the compact star. Alternatively, even a null detection of radio pulsations could still be helpful for providing some constraints on the binary properties, which would be beneficial for searching the pulses at X-rays or $\gamma$-rays which are less affected by wind absorption.

The paper is structured as follows. First, in Sect. 2, we describe the basic geometry of a pulsar orbiting around a massive star and the ingredients to calculate radio absorption. Then, in Sect. 3, we apply our model to two typical and well-studied binaries, namely, \PSRB\ and LS 5039. Finally, the conclusion and discussion of the implications on other binaries are presented in Sect. 4.

\section{Model description}
For HMGBs, the obscuration of radio pulsations depends on the density and temperature of the medium distributed along the LOS, which are mainly determined by their optical companions. A massive star can lose its mass via intense outflows, including a polar wind and an equatorial disc for some Oe/Be stars. In the isotropic case, the number density of the wind is given by (e.g. Waters et al. 1988)
\begin{eqnarray}\label{nwind}
  n_{\rm{w,i}}\left( r \right) &=& n_{\rm w,0}\left( \frac{r}{r_{\star}} \right) ^{-2},\label{nw}
\end{eqnarray}
where $r$ is the radial distance from the centre of the star. The base density $n_{\rm w,0}$ at the stellar surface $r_{\star}$ is related to the mass-loss rate and wind velocity as $n_{\rm w,0}={\dot{M}/4\pi r_{\star}^2v_{\rm w}\mu_{\rm i}m_{\rm p}}$, with $\mu_{\rm i}$ being the mean ion molecular weight and $m_{\rm p}$ being the mass of protons. Because of adiabatic cooling, the wind temperature decreases slowly with radial distance from the star. By ignoring the heating due to photoionisation, the temperature distribution of the wind can be expressed by the following power-law relation (Kochanek 1993; Bogomazov 2005):
\begin{eqnarray}\label{Twind}
  T_{\rm w}(r) &=& T_{\star}\left(\frac{r}{r_{\star}}\right)^{-\beta},
\end{eqnarray}
where $T_{\star}$ is the effective temperature of the star at $r_{\star}$. The value of the index $\beta$ depends on the adiabatic index of the wind gas and can be in the range of $\sim\left({2/3}-{4/3}\right)$. In the following calculations, we simply adopt $\beta=2/3$.

In addition to the stellar wind, some Oe/Be stars can further possess an equatorial decretion disc. The disc is believed to be responsible for the optical line emission and the infrared excess of these stars (see Porter \& Rivinius 2003 for a review on Be stars).
The density of the disc is usually assumed to follow a vertical Gaussian distribution as (e.g. Carciofi \& Bjorkman 2006)
\begin{eqnarray}\label{ndisc}
  n_{\rm{d,i}}\left( r_{\rm d}, z \right) &=& n_{\rm{d},0}\left( \frac{r_{\rm d}}{r_{\star}} \right) ^{-3.5}\exp \left[ -\frac{z^2}{2H^2(r_{\rm d})} \right],
\end{eqnarray}
where $r_{\rm d}$ is the radial distance along the mid-plane and $z$ is the vertical distance.
The scale height of the disc is given by
\begin{eqnarray}
  H(r_{\rm d}) &=& \frac{v_{\rm s}}{v_{\rm c}}\frac{r_{\rm d}^{3/2}}{r_{\star}^{1/2}},
\end{eqnarray}
where $v_{\rm s}=\sqrt{kT_{\star}/\mu m_{\rm p}}$ and $v_{\rm c}=\sqrt{GM_{\star}/r_{\star}}$ are the isothermal sound speed and the critical speed of the star, respectively. We note that $M_{\star}$ is the stellar mass and $\mu\simeq 0.62$ is adopted.
Assuming that the disc is vertically isothermal, the temperature profile with respect to the radial distance can be written as follows (e.g. Carciofi \& Bjorkman 2006):
\begin{eqnarray}\label{Tdisc}
  T_{\rm d}(r_{\rm d}) &=& \frac{T_{\star}}{\pi^{1/4}}\left[\arcsin\left(\frac{r_{\star}}{r_{\rm d}}\right)-
  \left(\frac{r_{\star}}{r_{\rm d}}\right)\sqrt{1-\left(\frac{r_{\star}}{r_{\rm d}}\right)^2}\right]^{1/4},
\end{eqnarray}
where the effects of viscous heating and irradiation from higher stellar latitudes on the disc are neglected.

\begin{figure}
  \centering
  \includegraphics[width=0.485\textwidth]{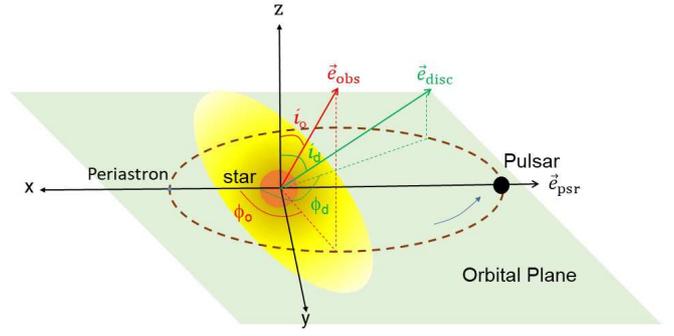}
  \caption{Illustration for the general case of the orbit with a pulsar orbiting around a massive star.} \label{fig:orbit}
\end{figure}

Since we are interested in the matter obstructing along the LOS that contributes to the dispersion measure (DM) and radio absorption, it is more convenient to express the above parameters in a proper coordinate system. As shown in Fig. \ref{fig:orbit}, with the origin of Cartesian coordinates located at the centre of the massive companion, we set the $x$-axis along the direction of the periastron and the $z$-axis perpendicular to the orbital plane. Then, the unit vectors in the direction of the orbital angular momentum, the pulsar, and the observer can be expressed as follows:
\begin{eqnarray}\label{vecs}
  \vec{e}_{\rm{orb}} &=& \left( 0,0,1 \right) \\
  \vec{e}_{\rm{psr}} &=& \left( \cos \phi ,\sin \phi ,0 \right) \\
  \vec{e}_{\rm{obs}} &=& \left( \sin i_{\rm{o}}\cos \phi _{\rm{o}},
  \sin i_{\rm{o}}\sin \phi _{\rm{o}}, \cos i_{\rm{o}} \right),
\end{eqnarray}
where $\phi$ is the true anomaly of the pulsar, $i_{\rm o}$ is the orbital inclination angle, and $\phi_{\rm o}$ is the true anomaly of the LOS projected on the orbital plane.

As shown in Fig. \ref{fig:B1259}, the unit vector of the stellar wind towards an arbitrary point $O$ along the LOS with a distance of $l$ from the pulsar is given by
\begin{eqnarray}\label{vec_wind}
  \vec{e}_{\rm{wind}} &=& \left(d\cdot\vec{e}_{\rm psr} + l\cdot\vec{e}_{\rm obs}\right)/r,
\end{eqnarray}
where $d=a(1-e^2)/(1+e\cos\phi)$ is the binary separation and $a$ and $e$ are the semi-major axis and the eccentricity of the orbit, respectively. The above equation can be rewritten as follows:
\begin{eqnarray}
  r^2 &=& d^2+l^2+2dl\left(\vec{e}_{\rm psr}\cdot\vec{e}_{\rm obs}\right).
\end{eqnarray}
The normal vector of the disc of the Oe/Be star is given by
\begin{eqnarray}\label{vec_disc}
  \vec{e}_{\rm{disc}} &=& \left( \sin i_{\rm{d}}\cos \phi _{\rm{d}}, \sin i_{\rm{d}}\sin \phi _{\rm{d}}, \cos i_{\rm{d}} \right),
\end{eqnarray}
where $i_{\rm d}$ is the polar angle measured from the normal direction of the orbit and $\phi_{\rm d}$ is the true anomaly of the disc normal projected on the orbital plane.
With the basic geometric relation, it is easy to obtain
\begin{eqnarray}
  r_{\rm d} &=& r\sin\vartheta,\   z=r\cos\vartheta\end{eqnarray}
with
\begin{eqnarray}
  \vartheta &=& \arccos(\vec{e}_{\rm wind}\cdot\vec{e}_{\rm disc}).
\end{eqnarray}
By substituting the above relationships into Eqs. (\ref{nwind}), (\ref{Twind}), (\ref{ndisc}), and (\ref{Tdisc}), we can obtain the density and temperature distributions of the stellar wind and disc along the LOS.

\begin{figure}
  \centering
  \includegraphics[width=0.485\textwidth]{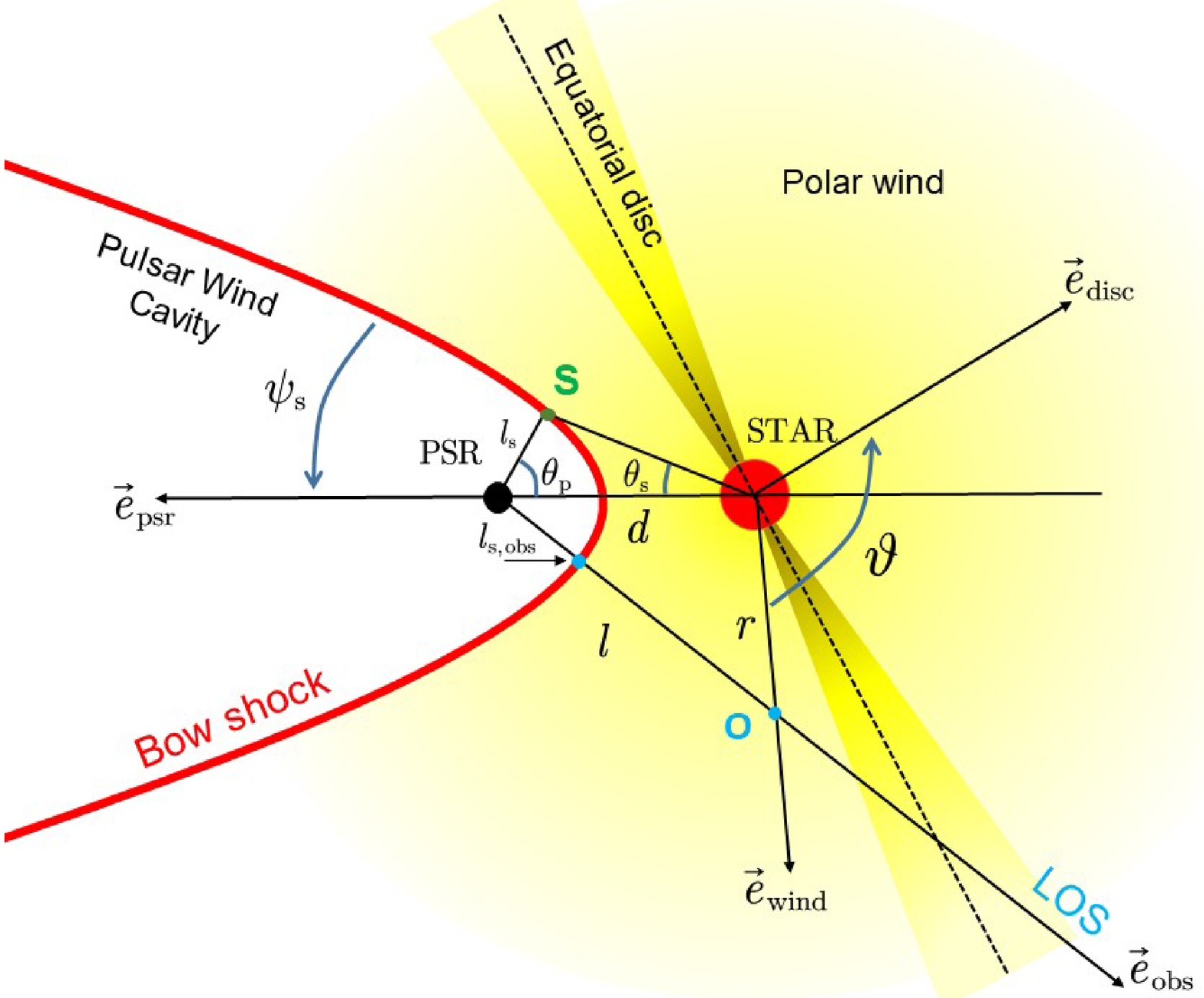}
  \caption{Edge-on viewing of the shock structure of \PSRB\ when the pulsar is moving around the periastron.} \label{fig:B1259}
\end{figure}

When the pulsar is moving around the massive star, the relativistic wind blown from the pulsar can create a cavity inside the stellar outflows. The cavity is shaped by a bow shock that arises from the dynamical balance between two winds, as shown in Fig. \ref{fig:B1259}. Denoting the distance of an arbitrary point $S$ on the shock contact discontinuity (CD) surface to the pulsar as $l_{\rm s}$, the shape of the surface can be described by the following geometric equations (Canto et al. 1996):
\begin{eqnarray}\label{shock}
  l_{\rm s} &=& d\sin\theta_{\rm s}\csc(\theta_{\rm s}+\theta_{\rm p})\end{eqnarray}
and
\begin{eqnarray}\label{angle}
  \theta_{\rm s}\cot\theta_{\rm s} &=& 1+\eta(\theta_{\rm p}\cot\theta_{\rm p}-1),
\end{eqnarray}
where $\theta_{\rm s}$ and $\theta_{\rm p}$ are the angles between the lines joining point $S$ on the CD surface to the pulsar and its companion related to the pulsar
vector. The momentum rate ratio of the two winds is given by
\begin{eqnarray}\label{eta}
  \eta &=& \frac{L_{\rm sd}/c}{\dot{M}v_{\rm w}},
\end{eqnarray}
with $L_{\rm sd}$ being the pulsar spin-down power and $c$ being the speed of light. For most HMGBs, the momentum rate of stellar outflows is usually much higher than that of pulsar wind (i.e. $\eta\ll1$), and then Eq. (\ref{angle}) can be approximated as follows:
\begin{eqnarray}\label{theta_s}
  \theta_{\rm s}&=&\left\{\frac{15}{2}\left[\sqrt{1+\frac{4}{5} \eta(1-\theta_{\rm p} \cot \theta_{\rm p})}-1\right]\right\}^{1 / 2}.\label{thetas}
\end{eqnarray}
When the LOS cuts through the shock cavity, the polar angle of the intersection is the following:
\begin{eqnarray}\label{theta_p}
  \theta_{\rm p,obs} &=& \pi-\arccos(\vec{e}_{\rm psr}\cdot\vec{e}_{\rm obs}).
\end{eqnarray}
In substituting Eqs. (\ref{theta_p}) and (\ref{theta_s}) into Eq. (\ref{shock}), it is easy to obtain the size of the cavity $l_{\rm s,obs}$ along the LOS. The asymptotic angle of the shock cavity can be approximated as (Eichler \& Usov 1993; Bogovalov et al. 2008)
\begin{eqnarray}
  \psi_{\rm s} &=& 2.1(1-\eta^{2/5}/4)\eta^{1/3},
\end{eqnarray}
by considering that $\theta_{\rm p}\sim (\pi-\theta_{\rm s})$ at infinity.

Finally, it is easy to obtain the orbital modulations of the DM and optical depth due to FFA by integrating the electron density and the absorption coefficient over the LOS:
\begin{eqnarray}
   DM &=& \int_{l_{\rm s,obs}}^{\infty} n_{\rm e} {\rm d}l,\label{DM} \\
   \tau(\nu) &=& \int_{l_{\rm s,obs}}^{\infty}{\alpha(\nu){\rm d}l}.\label{tff}
\end{eqnarray}
The contribution from the pulsar wind cavity is ignored here since the pair density of the pulsar wind is far less than that of the stellar outflows (Melatos et al. 1995). The FFA coefficient is given by (Rybicki \& Lightman 1979; Ghisellini 2013)
\begin{eqnarray}\label{aff}
  \alpha(\nu) &=& \frac{4}{3}\left( \frac{2\pi}{3} \right) ^{1/2}\frac{Z^2n_{\rm{e}}n_{\rm{i}}q_{\rm e}^{6}}{m_{\rm e}^{2}c^2}\left( \frac{m_{\rm e}c^2}{kT} \right) ^{1/2}\frac{1-e^{-h\nu /kT}}{h\nu ^3}\bar{g}_{\rm{ff}} \nonumber
  \\
   &\simeq& 0.018Z^2n_{\rm{e}}n_{\rm{i}}T^{-3/2}\nu ^{-2}\bar{g}_{\rm{ff}},
\end{eqnarray}
where
\begin{eqnarray}
  \bar{g}_{\rm{ff}} &\simeq& \frac{\sqrt{3}}{\pi}\ln\left[\frac{(kT)^{3/2}}{2\pi\nu Zq_{\rm e}^2m_{\rm e}^{1/2}}\right]
\end{eqnarray}
is the Gaunt factor, $Z^2\simeq1.4$, and $T$ is the absorbing medium temperature. The second expression of Eq. (\ref{aff}) is obtained under the Raleigh-Jeans approximation ($h\nu\ll kT$). The number densities of electrons and ions are related by $n_{\rm e}=n_{\rm i}\mu_{\rm i}/\mu_{\rm e}$, with $\mu_{\rm i}=4/(1+3X)\simeq1.29, \mu_{\rm e}=2/(1+X)\simeq1.18$ for the typical values of stellar outflows, and $X\sim0.7$ is the hydrogen abundance (Zdziarski et al. 2010; Dubus 2013).

It is necessary to point out that in the above descriptions, the shock is assumed to have an asymptotic radial structure that is directing away from the massive star, and the effect of orbital motion on the shock structure is not considered.
However, for HMGBs, in particular for LS 5039 which has a more compact orbit, the putative pulsar's fast orbital motion would significantly deflect the shock structure (Bosch-Ramon \& Barkov 2011; Zabalza et al. 2013; Bosch-Ramon 2021). Furthermore, due to the Coriolis force induced by orbital motion, the pulsar wind would also be terminated by an additional shock located in the opposite direction of the star. The presence of this Coriolis shock was predicted in Bosch-Ramon \& Barkov (2011) and later confirmed by numerical simulations (Bosch-Ramon et al. 2012, 2015; Huber et al. 2021a,b). We explore and discuss the influence of this Coriolis shock in the case of LS 5039 in a later section.
As for PSR B1259-63/LS 2883, due to its relatively larger binary separation, the Coriolis shock is located far from the companion star, where the density of the stellar outflows is very low. Furthermore, because of the small orbital inclination angle (i.e. $180^{\circ}-i_{\rm o}\sim 26^{\circ}$), the LOS will not directly point to this Coriolis shock. Therefore, the effect of the Coriolis shock on the DM and radio absorption of PSR B1259-63 can be ignored.

\section{Results}
Among the detected HMGBs, \PSRB\ and \LS\ are the most well-studied binaries in both observational and theoretical aspects. In this section, we use the model described in Sect. 2 to investigate the orbital modulations of the DM and radio absorption of these two systems. The observational parameters are listed in Table. 1.

\begin{table*}[t]
\caption{Observational parameters of \PSRB \ and \LS. \label{table}}
\begin{tabular}{l c c c c}
\hline
\textbf{Parameter}   & \textbf{\PSRB}  & \textbf{Reference} & \textbf{\LS}  & \textbf{Reference}\\
\hline
eccentricity, $e$  &   0.86987970  & (1)& 0.35  & (4,5)\\
semi-major axis, $a$ (\rm {AU}) & 5.9257966 &(1)& 0.14  & (4)\\
orbital period, $P_{\textrm{orb}}$ (\textrm{days})    & 1236.724526 & (1)& 3.906  & (4,5,6)\\
\hline
the inclination angle of the observer, $i_{\rm o}$   & $154^{\circ}$ & (2)& $13-64^{\circ}$  & (4)\\
the true anomaly of the observer, $\phi_{\rm o}$   & $132^{\circ}$  & (2)& $224^{\circ}$  & (4)\\
\hline
stellar mass, $M_\star\ (M_{\odot})$   & $30$  & (3)& 22.9  & (4)\\
stellar radius, $r_\star\ (R_{\odot})$   & $9.2$ & (3)& 9.3  & (4)\\
stellar temperature, $T_\star$ (\textrm{K})  & $3.02\times10^4$  & (3)&  $3.90\times10^4$  & (4)\\
\hline
\end{tabular}
\\
{{
\textbf{Reference.} (1) Shannon et al. (2014); (2) Miller-Jones et al. (2018); (3) Negueruela et al. (2011); (4) Casares et al. (2005); (5) Aragona et al. (2009); (6) Sarty et al. (2011).
}}
\end{table*}

\subsection{PSR B1259-63/LS 2883}

\begin{figure}
  \centering
  \includegraphics[width=0.485\textwidth]{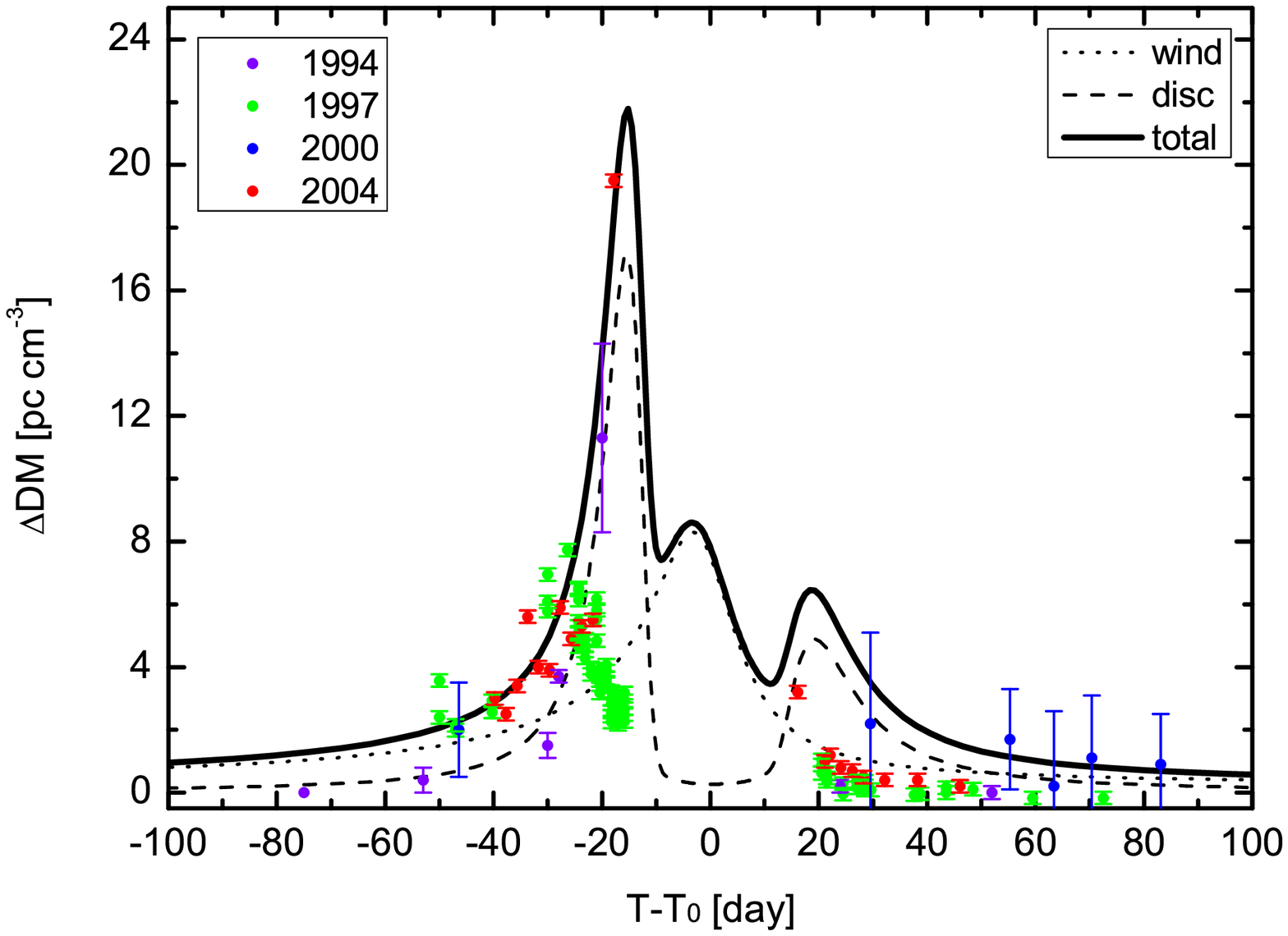}
  \includegraphics[width=0.485\textwidth]{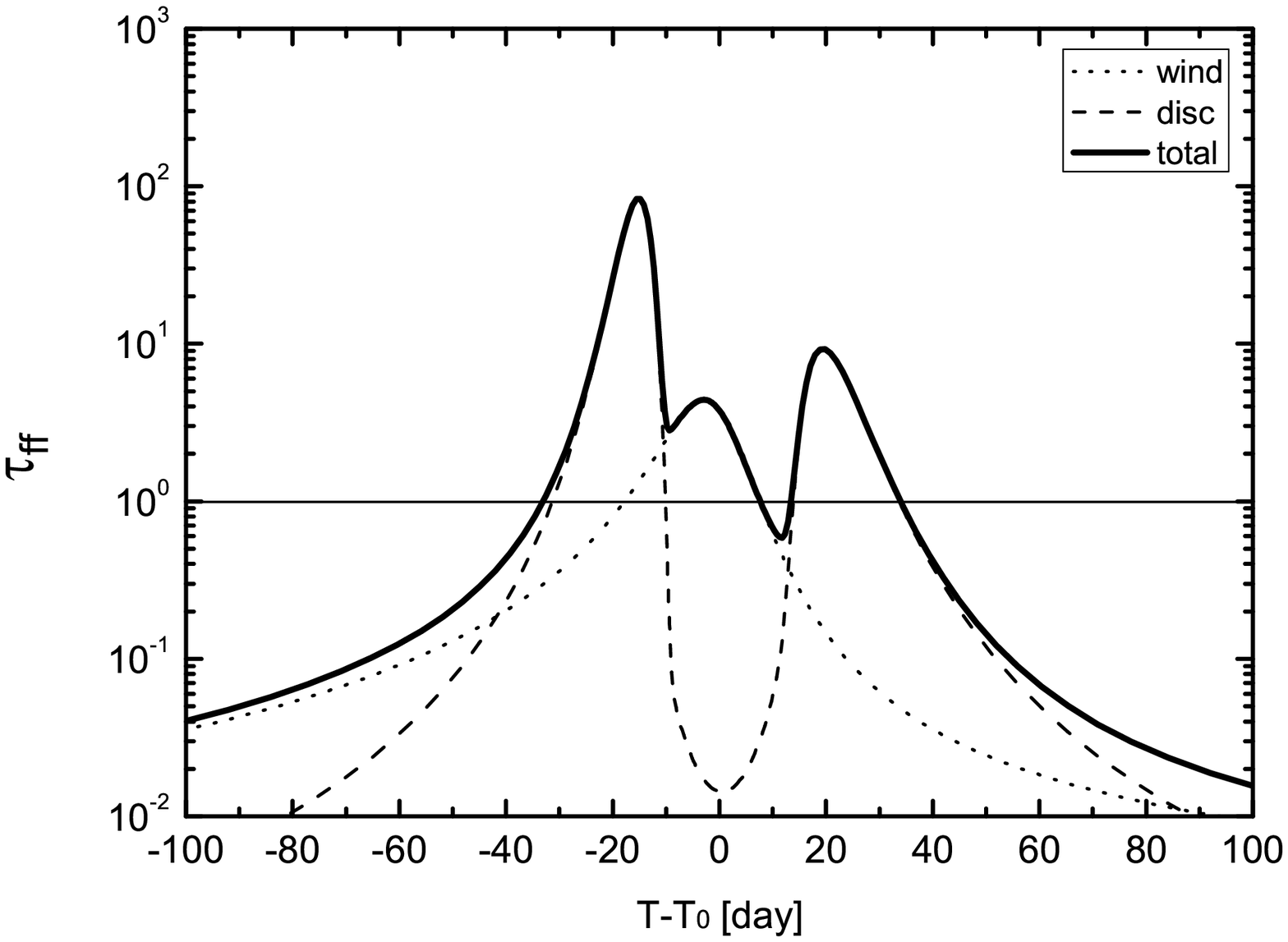}
  \caption{DM variations (upper panel) and the optical depth at $\nu=1.4\ {\rm GHz}$ (bottom panel) for PSR B1259-63. The dotted lines and dashed lines correspond to the contributions from the polar wind and equatorial disc, respectively. The solid lines are the sum of both components.
  The wind parameters adopted in the calculations are $\dot{M}=2\times10^{-8}M_{\odot}/{\rm yr}^{-1}$, $v_{\rm w}=3\times10^8\ {\rm cm/s}$, therefore, the corresponding wind base density is $n_{\rm w,0}\simeq4.12\times10^8{\rm cm^{-3}}$, and the momentum rate ratio is $\eta\simeq0.07$. The disc parameters adopted here are $n_{\rm d,0}=4.5\times10^{11}\ {\rm cm^{-3}}$, $i_{\rm d}=60^{\circ}$, and $\phi_{\rm d}=188^{\circ}$, which are obtained by fitting the DM data.
  The data points in the upper panel are the observed DM variations of PSR B1259-63 during the 1994, 1997, 2000, and 2004 periastron passages which are taken from Johnston et al. (1996), Wang et al. (2004), Connors et al. (2000), and Johnston et al. (2005), respectively.} \label{fig:B1259_fitting}
\end{figure}

\begin{figure}
  \centering
  \includegraphics[width=0.485\textwidth]{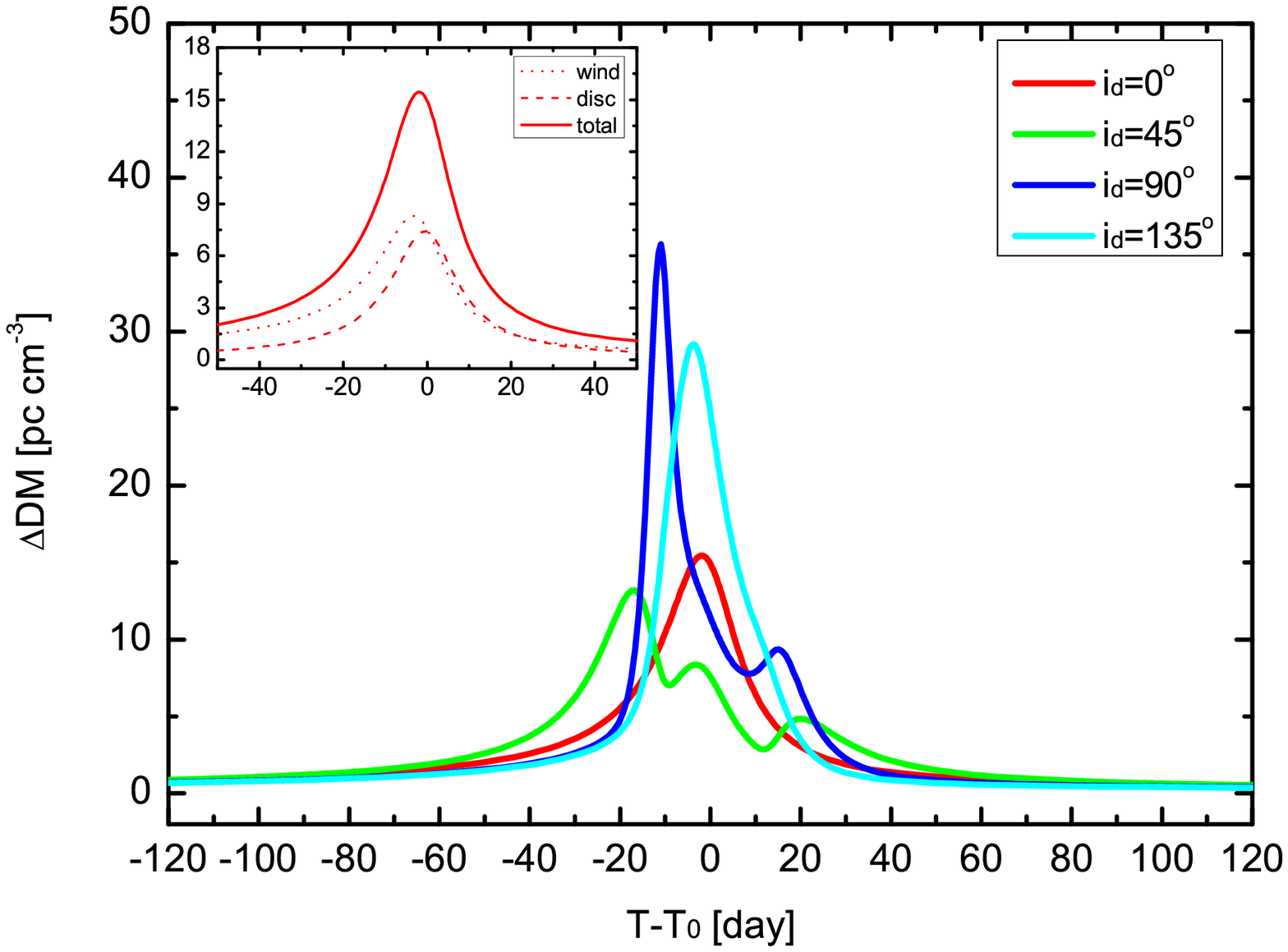}
  \includegraphics[width=0.485\textwidth]{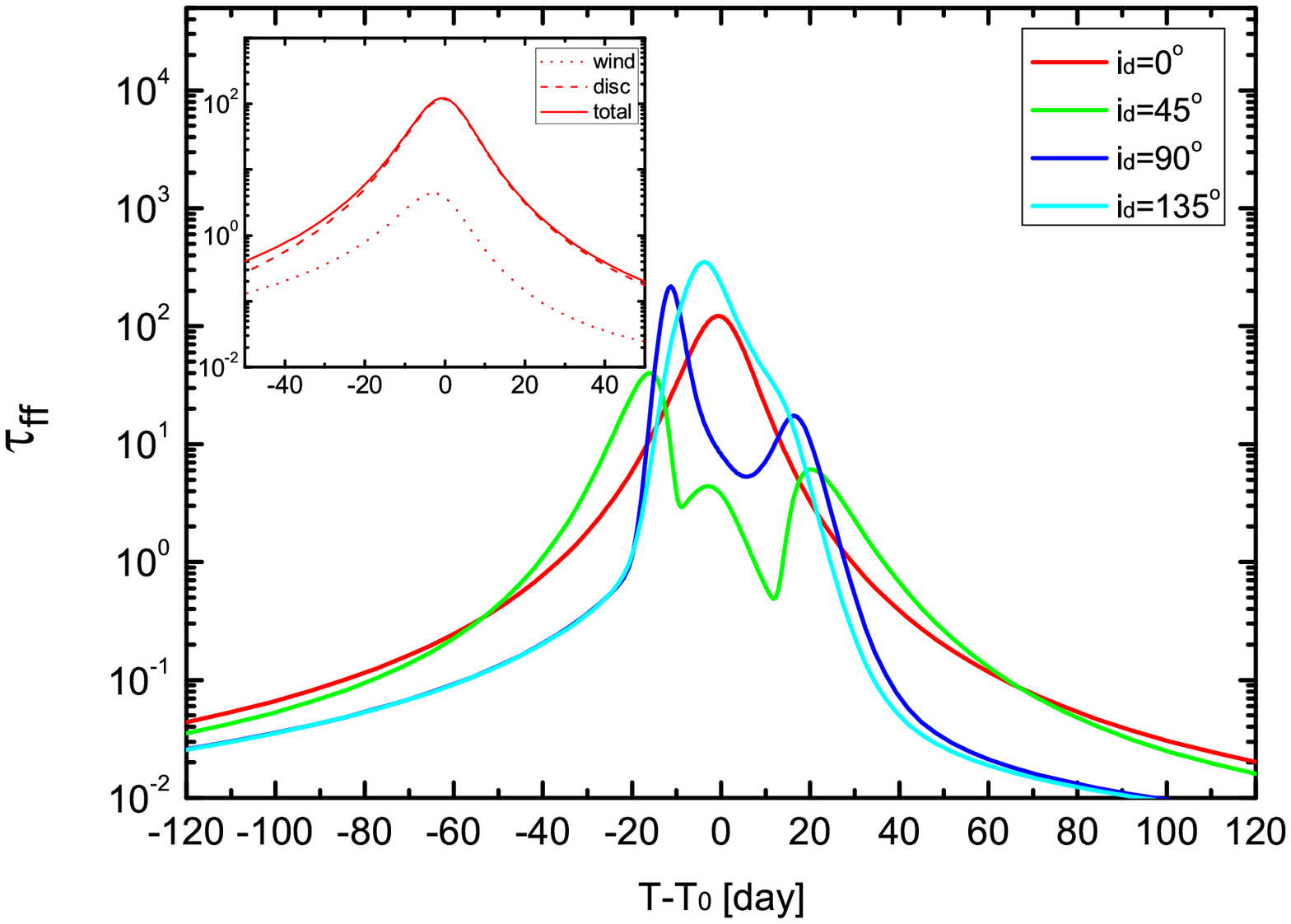}
  \caption{Same as Fig. \ref{fig:B1259_fitting}, but for different disc inclination angles ($i_{\rm d}=0^{\circ}, 45^{\circ}, 90^{\circ}$, and $135^{\circ}$). The inset plots show the case in which the disc is completely immersed in the orbital plane (i.e. $i_{\rm d}=0^{\circ}$) with the dotted lines and dashed lines corresponding to the contributions from the stellar wind and equatorial disc, respectively. The solid lines are the sum of both components.} \label{fig:B1259_disc}
\end{figure}

PSR B1259-63 is the first radio pulsar discovered in orbit around a massive star LS 2883, with an eccentricity of $e\simeq0.87$ and an orbital period of $P_{\rm orb}\simeq 1236.7$ days (Johnston et al. 1992; Shannon et al. 2014). The spin period of PSR B1259-63 is 47.76 ms with the spin-down luminosity of $L_{\rm sd}\simeq8.2\times10^{35}\ {\rm erg/s}$. The companion star is a type O9.5Ve star, with stellar outflows containing a dilute polar wind and a dense equatorial disc inclined to the orbital plane (Negueruela et al. 2011).

The radio pulsation of PSR B1259-63 vanishes around the periastron caused by the absorption of the intense outflows from its massive companion.
In Fig. \ref{fig:B1259_fitting}, we present the orbital modulations of the DM and optical depth of PSR B1259-63 due to FFA at the frequency of $\nu=1.4\ {\rm GHz}$. The curves are calculated with the following parameters: the stellar mass-loss rate $\dot{M}=2\times10^{-8}M_{\odot}/{\rm yr}^{-1}$ and the wind velocity $v_{\rm w}=3\times10^8\ {\rm cm/s}$, and therefore the corresponding momentum rate ratio is $\eta\simeq0.07$. The base number density of the wind and the disc are $n_{\rm w,0}\simeq4.12\times10^8\ {\rm cm^{-3}}$
and $n_{\rm d,0}=4.5\times10^{11}\ {\rm cm^{-3}}$, respectively.
The corresponding disc base density ($\rho_{\rm d,0}\simeq8.36\times10^{-13}\ {\rm g\cdot cm^{-3}}$) is in the range of typical values as suggested in previous studies for Be stars (e.g. Porter 2003; McSwain et al. 2008; Carciofi \& Bjorkman 2008). For the disc orientation, we use the result of Chen et al. (2019), which was obtained by fitting the multi-wavelength light curves with $\phi_{\rm d}=188^{\circ}$ and a relatively larger inclination angle of $i_{\rm d}=60^{\circ}$. The opening angle of the disc projected on the orbital plane can be calculated by
\begin{eqnarray}
  \Delta\phi_{\rm d} &=& \arcsin\left(\frac{\sin\Delta\theta_{\rm d}}{\sin i_{\rm d}}\right)
,\end{eqnarray}
with $\Delta\theta_{\rm d} = \arcsin\left[H(r_{\rm d})/r_{\rm d}\right]$. By substituting $r_{\rm d}\sim d$ during the disc passages, it is easy to obtain $\Delta\phi_{\rm d}\simeq10.5^{\circ}$}. It is worth noting that the momentum rate of the disc is much higher than that of the polar wind, and the additional pressure from the stellar disc would push the shock much closer to the pulsar during the disc passages (e.g. Chen et al. 2019). Therefore, when the pulsar is moving inside the disc, we assume that the momentum rate ratio is reduced by two orders of magnitude (i.e. $\sim 0.01\eta$, where $\eta$ is given by Eq. \ref{eta}). This would result in the size of the shock cavity along the LOS to shrink by an order of magnitude during the disc passages.
In the figures, the dotted lines and dashed lines are the contributions of the stellar wind and the equatorial disc, respectively. The solid lines are the sum of both components.
A noticeable feature of the curves is the asymmetric double-hump structure during the disc passages, with an additional peak around the periastron corresponding to the contribution from the polar wind. As shown via the dashed lines, the first peak is significantly higher than
that after the periastron since the disc is tilted with respect to the major axis of the orbit (i.e. $\phi_{\rm d}\neq 0^{\circ}$ or $180^{\circ}$) and the local density encountered by the pulsar would be more prominent during the disc passage before the periastron. Furthermore, the pulsar is located at the far side of the companion with respect to the observer before the periastron and, therefore, the path length along the LOS is larger, resulting in a higher DM and optical depth.
As the pulsar moves around the periastron, the LOS crosses the outermost edge of the disc. Therefore, there would be only a tiny contribution from the equatorial disc, and the DM and absorption are dominated by polar wind.
The calculated DM curve is roughly consistent with the observed data except for the 1997 periastron passage, where the DM value started to decline at $\sim30$ days before the periastron.
The discrepancies of the observed DM data from orbit to orbit could be caused by the changes of the medium around the pulsar, including the pulsations or destructions of the disc and the inhomogeneities of the stellar outflows.

Given the significant importance of the disc, we investigate the influences of the disc inclination angles on the results, as presented in Fig. \ref{fig:B1259_disc}. The inset plots show the case in which the disc is completely immersed in the orbital plane (i.e. $i_{\rm d}=0^{\circ}$). In this case, the DM and absorption are dominated by the equatorial disc, and they reach their maxima at periastron.
As $i_{\rm d}$ increases, the two-peak structures appear in the DM and radio optical depth curves, and the LOS starts to approach the inner part of the disc with a higher density as the pulsar moves around periastron. When $i_{\rm d}\geq90^{\circ}$, the pulsar is eclipsed by the dense disc around periastron, and the radio pulsations are again completely absorbed.

\subsection{\LS}

\begin{figure}
  \centering
  \includegraphics[width=0.485\textwidth]{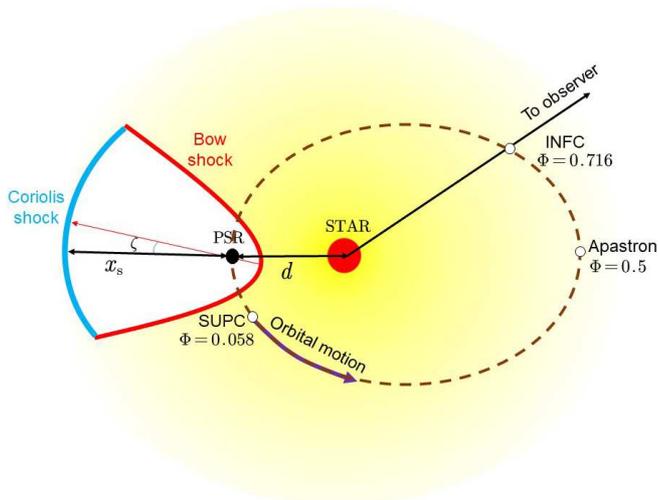}
  \caption{Simple illustration of the shock structure for LS 5039 as seen from above the orbital plane.} \label{fig:LS5039_shock}
\end{figure}

\begin{figure}
  \centering
  \includegraphics[width=0.485\textwidth]{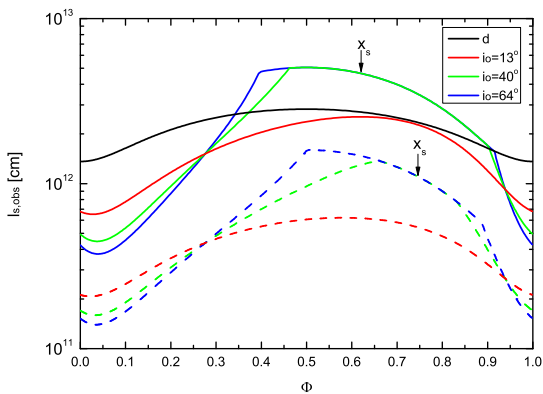}
  \caption{Orbital modulations of the cavity size along the LOS, $l_{\rm s,obs}$, for LS 5039 with different orbital inclination angles. The adopted pulsar spin-down luminosity and the stellar wind velocity are $L_{\rm sd}=6\times10^{36}\ {\rm erg/s}$ and $v_{\rm w}=3\times10^8\ {\rm cm/s}$. Solid lines represent the cases with $\dot{M}=1\times10^{-7}\ {M_{\odot}/{\rm yr}}$ (i.e. the corresponding momentum rate ratio is $\eta\simeq0.1$), while dashed lines are for $\dot{M}=1\times10^{-6}\ {M_{\odot}/{\rm yr}}$ (i.e. $\eta\simeq0.01$). The black solid line shows binary separation for comparison.} \label{fig:LS5039_ls}
\end{figure}

In contrast to \PSRB, LS 5039 has a relatively lower eccentricity ($e\sim 0.35$) with a much tighter orbit ($a\sim 0.14$ AU). The short orbital period ($P_{\rm orb}\simeq3.9\ {\rm days}$) makes \LS\ a perfect target to investigate its multi-wavelength emissions along the entire orbit. The optical companion of LS 5039 is identified as a type O6.5V massive star, while the nature of the compact object is still unknown (McSwain et al. 2001, 2004; Casares et al. 2005; Aragona et al. 2009; Sarty et al. 2011). The inclination angle of the orbit is not well constrained, ranging from $13^{\circ}$ to $64^{\circ}$ (Casares et al. 2005). According to the binary mass function, a relatively low inclination angle ($i_{\rm o}\leq 30^{\circ}$) favours a black hole primary, while a higher value suggests a neutron star origin.

Radio observations of LS 5039 have discovered a persistent emission, but they have not found any pulsation. The radio emission shows an extended morphology, which was initially attributed to the structure of a micro-quasar jet (Rib{\'o} et al. 1999; Paredes et al. 2000, 2002).
However, Rib{\'o} et al. (2008) found that the changes in radio morphology may not be consistent with the micro-quasar scenario. Instead, the pulsar wind interaction scenario seems to be favoured since the wind interaction region could also exhibit a comet-like tail that mimics a micro-quasar jet (Mold{\'o}n et al. 2012; Marcote et al. 2015). The wind interaction scenario is also supported by dynamic simulations and the multi-wavelength modelling of LS 5039 (e.g. Dubus et al. 2015; Molina \& Bosch-Ramon 2020; Huber et al. 2021b). If the compact object in LS 5039 is indeed a rotation-powered pulsar, the non-detection of radio pulsations could be caused by severe absorption by the intense stellar outflow, in particular considering the tight orbit of this binary.

Due to the compact orbit, the wind interaction region of LS 5039 can be modified by the Coriolis force induced by the fast motion of the presumed pulsar. First, the bow shock close to the pulsar is deflected by an angle of $\zeta\simeq \arctan(v_{\rm p}/v_{\rm w})$, where $v_{\rm p}$ is the orbital velocity (e.g. Zabalza et al. 2013). For LS 5039, the deflection angle is of the order of $\zeta\sim5-12^{\circ}$ along the orbit. The angle is small so we assume that the shock still remains as a radial structure that is directed away from the star for simplicity.
Second, at a larger distance, the shocked flow would be disturbed due to fluid instabilities and wind mixing, and a Coriolis shock would be formed in the opposite direction from the star (Bosch-Ramon \& Barkov 2011; Bosch-Ramon et al. 2012, 2015; Huber et al. 2021a,b). The real structure of the interaction region could be very complicated, and it is not easy to be modelled explicitly in a simple analytical way. Therefore, we use a simplified structure containing the bow shock and the Coriolis shock, as adopted in previous studies (e.g. Zabalza et al. 2013; Takata et al. 2014). In Fig. \ref{fig:LS5039_shock}, we show the illustration of the shock structure for LS 5039 as seen from above the orbital plane, where the periastron is set at $\Phi=0$, while the superior and inferior conjunctions are at $\Phi=0.058$ and 0.716, respectively (Casares et al. 2005). The position of the Coriolis shock is estimated by the dynamic balance between the ram pressure of the stellar wind due to the Coriolis force and that of the pulsar wind, which yields (Bosch-Ramon \& Barkov 2011)
\begin{eqnarray}\label{shock_cor}
  x_{\rm s} &\simeq&  \left(\frac{L_{\rm sd}v_{\rm w}}{\dot{M}c(2\Omega)^2}\right)^{1/2},
\end{eqnarray}
where $\Omega$ is the orbital angular velocity. As the LOS is located inside the bow shock cavity, the size of the shock cavity $l_{\rm s,obs}$ given by Eq. (\ref{shock}) should be replaced with Eq. (\ref{shock_cor}).
The orbital modulations of the cavity size along the LOS with different orbital inclination angles are shown in Fig. \ref{fig:LS5039_ls}. For a modest orbital inclination angle $i_{\rm o}=40^{\circ}$ and momentum rate ratio $\eta\simeq0.1$ (the corresponding half-opening angle of the bow shock is $\psi_{\rm s}\simeq51^{\circ}$), the LOS would point to the Coriolis shock region during nearly half of the orbit. This is because the solid angle of the Coriolis shock measured from the pulsar is larger than that measured from the star. Obviously, for a lower value of the momentum rate ratio (e.g. $\eta\simeq0.01$), the size of the cavity along the LOS would get smaller.

\begin{figure}
  \centering
  \includegraphics[width=0.485\textwidth]{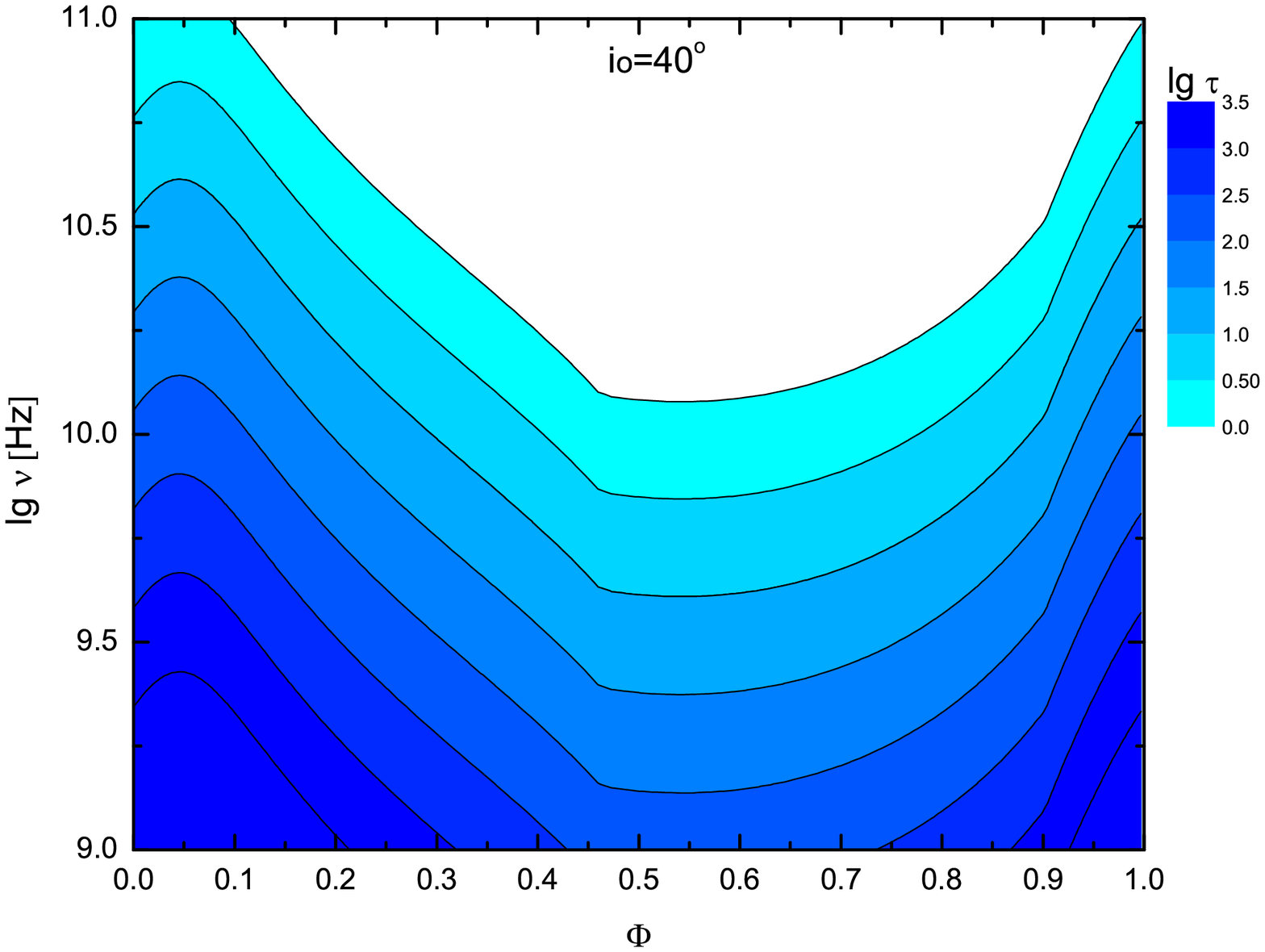}
  \includegraphics[width=0.485\textwidth]{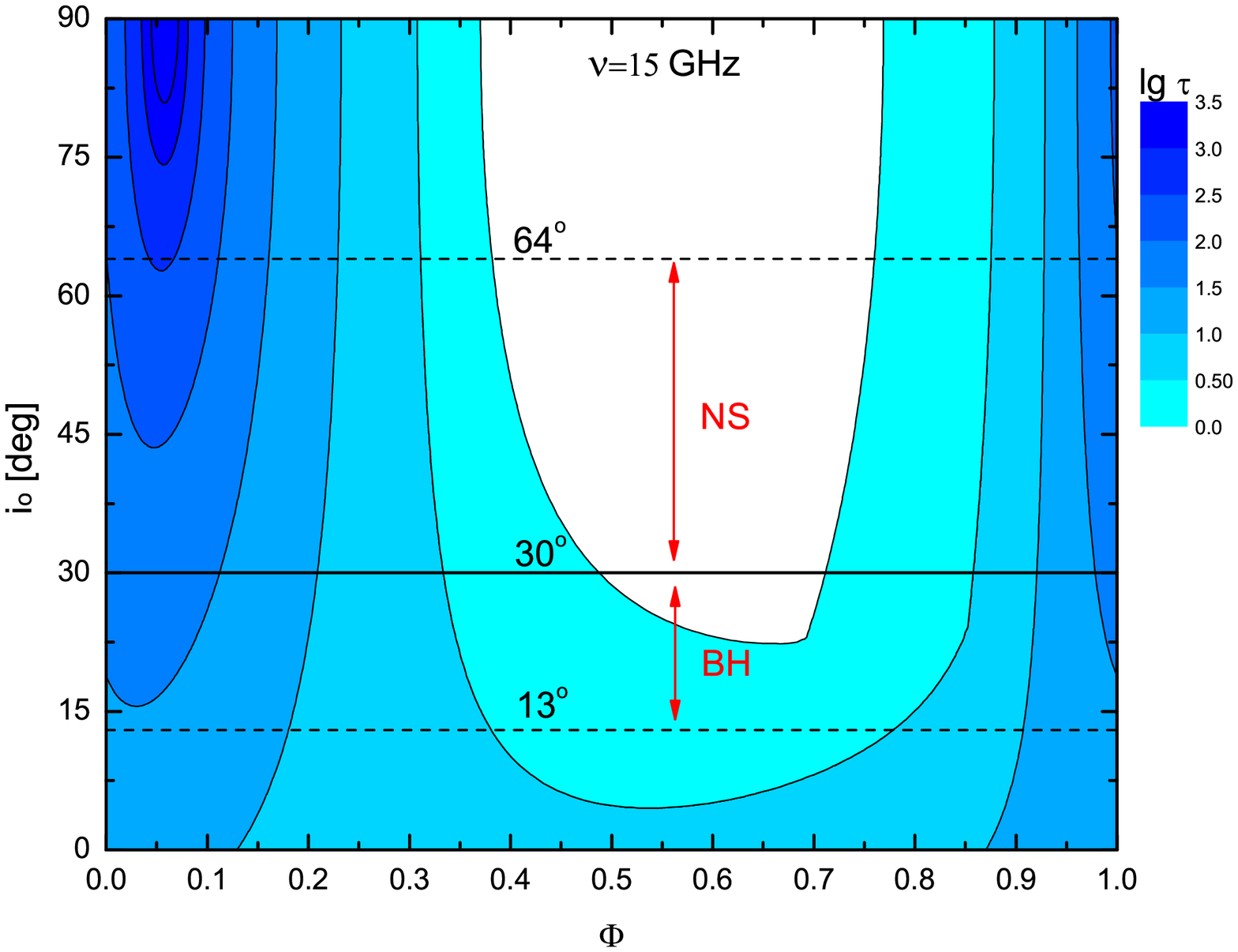}
  \caption{Orbital modulations of the optical depth for LS 5039 at different frequencies (upper panel) and orbital inclination angles (bottom panel). According to the binary mass function, $13^{\circ}< i_{\rm o}< 30^{\circ}$ suggests that the compact object is a black hole, while a higher value (i.e. $30^{\circ}< i_{\rm o}< 64^{\circ}$) favours a neutron star primary. The model parameters adopted here are the same as those in Fig. \ref{fig:LS5039_ls} with $\dot{M}=1\times10^{-7}\ {M_{\odot}/{\rm yr}}$. } \label{fig:LS5039_tau1}
\end{figure}

\begin{figure}
  \centering
  \includegraphics[width=0.485\textwidth]{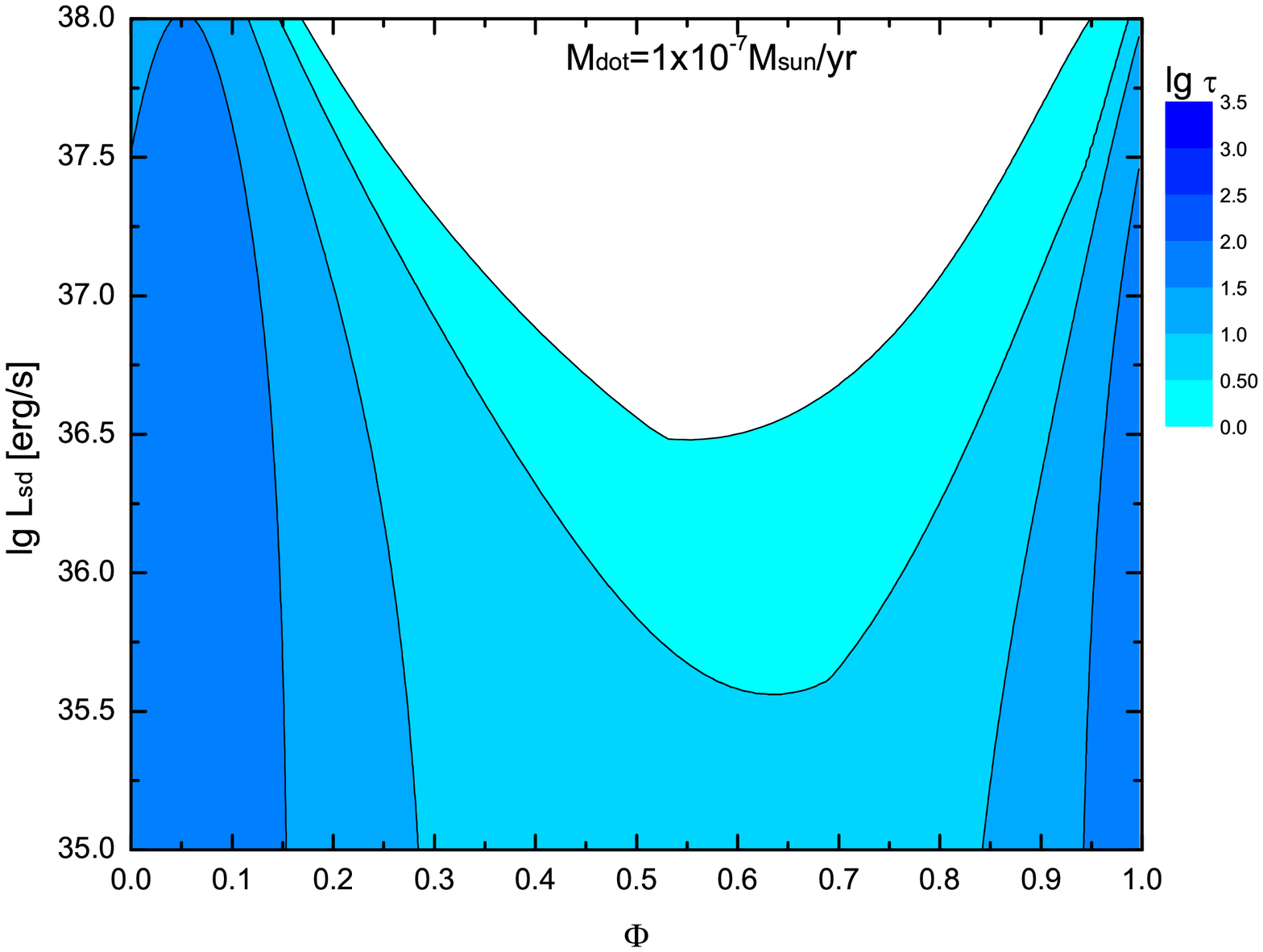}
  \includegraphics[width=0.485\textwidth]{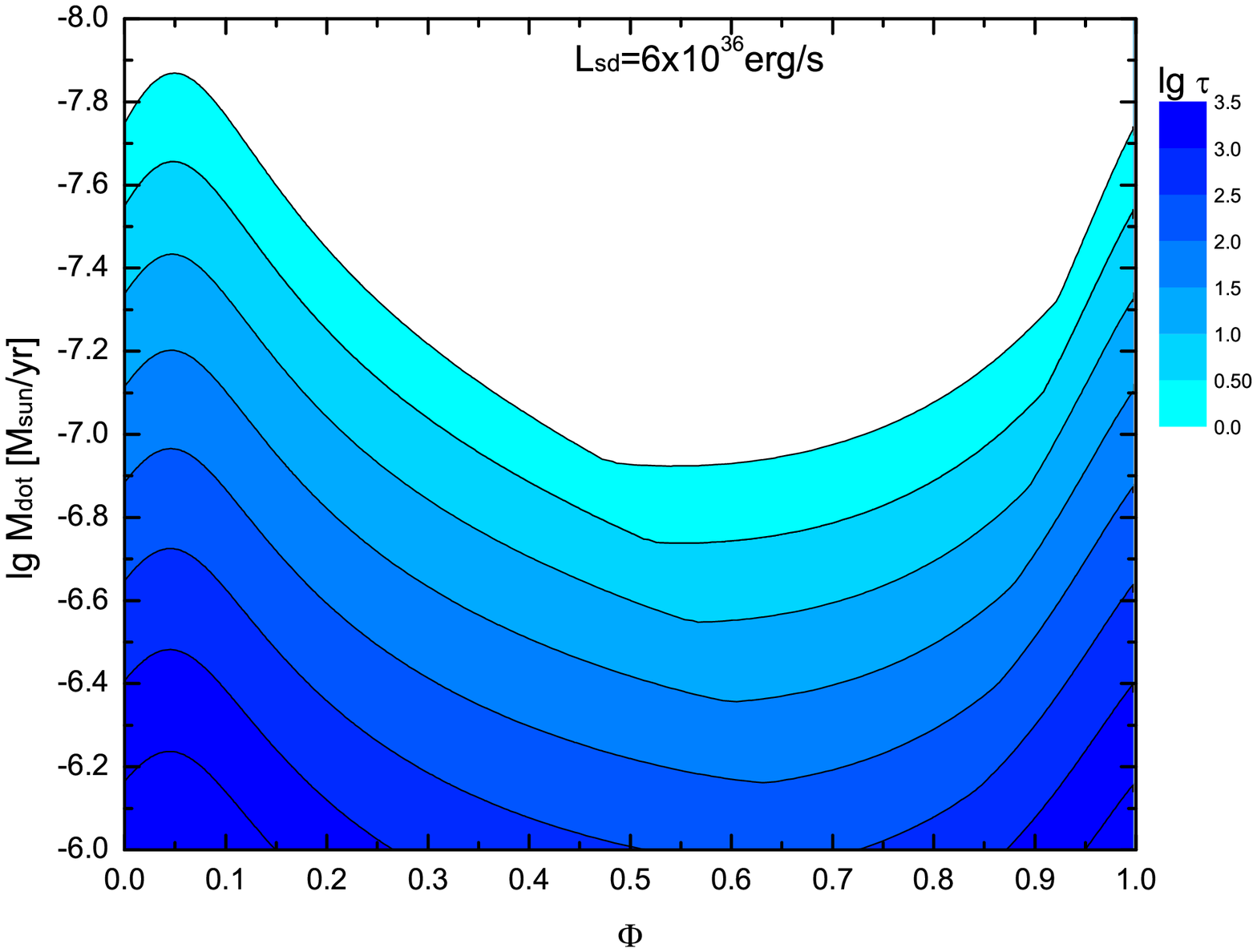}
  \caption{Orbital modulations of the optical depth for LS 5039 at the frequency of $\nu=15\ {\rm GHz}$ with different spin-down powers (upper panel) and stellar mass-loss rates (bottom panel). The orbital inclination angle is fixed as $i_{\rm o}=40^{\circ}$. } \label{fig:LS5039_tau2}
\end{figure}

For LS 5039, many parameters remain unknown as of yet, including the orbital inclination angle, the stellar mass-loss rate, and the spin-down luminosity of the presumptive pulsar. In Fig. \ref{fig:LS5039_tau1} and \ref{fig:LS5039_tau2}, we explore the dependence of absorption on the observing frequencies and the above parameters.
With the typical values of binary parameters, the stellar wind could always be opaque along the orbit if the observed frequency is below $10\ {\rm GHz}$ as shown in the upper panel of Fig. \ref{fig:LS5039_tau1}. Since the optical depth due to FFA decreases as $\tau\propto\nu^{-2.1}$, a higher frequency is preferable to minimise absorption. If the putative pulsar in LS 5039 is similar to PSR B1259-63 with pulsed emission extending above $10\ {\rm GHz}$ (e.g. Wang et al. 2005), it is still possible to detect the radio pulsations. The bottom panel of Fig. \ref{fig:LS5039_tau1} shows the orbital modulations of the optical depth at the frequency of $\nu=15\ {\rm GHz}$ with different orbital inclination angles. It can be found that, with the increase in $i_{\rm o}$, the amplitude of the optical depth variation becomes more significant, and there could be a short transparent window for radio pulsation around the INFC as long as the orbit is appropriately inclined.

For HMGBs, the structure of the shock cavity is governed by the momentum rate ratio of the two winds, which is mainly determined by the pulsar spin-down luminosity and the stellar mass-loss rate. Therefore, in Fig. \ref{fig:LS5039_tau2}, we investigate the effects of $L_{\rm sd}$ and $\dot{M}$ on the optical depth at the frequency of $\nu=15\ {\rm GHz}$. Obviously, a higher spin-down luminosity means that the pulsar would create a larger cavity inside the stellar wind and, therefore, the optical depth would be smaller. The high $\gamma$-ray luminosity of LS 5039 implies that its compact object might be an energetic pulsar with the spin-down power larger than $10^{36}\ {\rm erg/s}$ (e.g. Bosch-Ramon 2021).
The stellar mass-loss rate can be derived from the H$\alpha$ profile. For LS 5039, $\dot{M}$ is not well determined yet, ranging from $5\times10^{-8}\ {M_{\odot}/{\rm yr}}$ to $7.5\times10^{-7}\ {M_{\odot}/{\rm yr}}$ (e.g. McSwain et al. 2004; Casares et al. 2005; Bosch-Ramon et al. 2007; Bosch-Ramon 2010; Szsostek \& Dubus 2011). In the bottom panel of Fig. \ref{fig:LS5039_tau2}, we cover the possible range of $\dot{M}$ suggested by previous studies. Compared to the pulsar spin-down power, the radio optical depth is more sensitive to the mass-loss rate since it is the main factor determining the size of the shock cavity and the wind density along the LOS.

\begin{figure}
  \centering
  \includegraphics[width=0.485\textwidth]{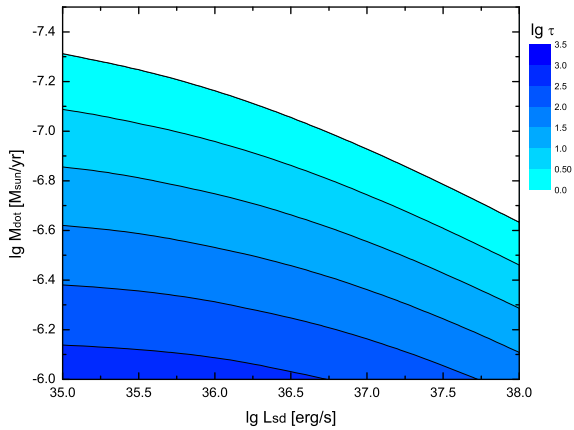}
  \caption{Dependence of the optical depth at the frequency of $\nu=15\ {\rm GHz}$ on the pulsar spin-down luminosity and stellar mass-loss rate when the pulsar is at the INFC. The orbital inclination angle $i_{\rm o}\simeq40^{\circ}$ is adopted in the calculations.} \label{fig:LS5039_tau3}
\end{figure}

In summary, if the compact object of LS 5039 is a rotation-powered pulsar, which is similar to PSR B1259-63, there could still be a transparent phase in which we can detect its radio pulsations even though it suffers from strong absorption
by the intense stellar wind along most of the orbit. The existence of this transparent window requires a high pulsar-to-stellar wind momentum rate ratio and an appropriate inclination angle of the orbit.
On the other hand, the non-detection of radio pulsations can still give further constraints on the system parameters.
McSwain et al. (2011) conducted searches of pulsed radio signals of LS 5039 with the Green Bank Telescope using S-band (1.7-2.4 GHz) and C-band (4.4-5.2 GHz) receivers, covering the orbital phases of 0.56-0.62 and 0.47-0.48, respectively. Both of which concluded with the null detection of pulsations. In substituting the above frequencies and the corresponding wind parameters along the LOS at these epochs into $\tau\geq1$, it is easy to obtain the momentum rate ratio with $\eta\leq0.41$ and $\eta\leq0.22$, with the orbital inclination angle being fixed with $i_{\rm o}= 40^{\circ}$. Observationally, the inclination angle of the orbit and the mass-loss rate of stars can be measured via optical spectroscopy. Therefore, the pulsar spin-down luminosity can be constrained.
In Fig. \ref{fig:LS5039_tau3}, we show the dependence of the optical depth at $\nu=15\ {\rm GHz}$ on the spin-down luminosity and stellar mass-loss rate for LS 5039 when the pulsar is moving around the INFC. With a typical value of $\dot{M}\sim 10^{-7} M_{\odot}/{\rm yr}$ for O stars, the null detection of radio pulsations at $\nu=15\ {\rm GHz}$ around INFC would imply that the pulsar spin-down power should be less than $5\times10^{36}\ {\rm erg/s}$.

\section{Discussion and conclusion}
HMGBs are unique laboratories for us to study the evolution of binaries and the properties of massive stars. The lack of accretion signal detection in HMGBs suggests that the compact objects may be non-accreting neutron stars, even though most of them are not confirmed.
In this paper, we have described the basic geometry and formulae to calculate the dispersion measure and radio absorption of HMGBs containing a pulsar in binary with a massive star, and we have applied our model to two typical HMGBs.
For the PSR B1259-63 system with a decretion disc around its massive companions, the radio pulsation is less affected by the polar wind due to its large binary separation, and the absorption is usually dominated by inclined discs. The position and orientation of the disc can be obtained by the obscuration of radio pulsation or the multi-wavelength fitting of light curves (e.g. Chen et al. 2019). We studied the effects of the inclined disc of LS 2883 on the radio pulsation from PSR B1259-63. The observed DM data are consistent with the equatorial disc inserted on the orbital plane with an inclination angle of $i_{\rm d}\sim60^{\circ}$ and the true anomaly of the disc normal projected on the orbital plane is $\phi_{\rm d}\sim188^{\circ}$. The corresponding opening angle of the disc projected on the orbital plane is about $10^{\circ}$. Our model for PSR B1259-63/LS 2883 can be easily applied to other similar Be/gamma-ray binaries, such as \PSRJ, \HESSJ,\ and \LSI. In particular, the last two have compact objects with an unknown nature as of yet.
For LS 5039-like systems, due to their compact orbits, intense outflows from their massive companions make detecting radio pulsations unlikely. Several tentative searches for radio pulsations turn out as null detections even at the apastron. We have shown that the shock cavity formed by wind collision and orbital motion could shield the putative pulsar from the intense outflows. We studied the effects of the orbital inclination angle and wind parameters on the radio absorption process. We have shown that there could be a transparent window at higher frequencies to detect radio pulsations if the momentum rate ratio of two winds is large enough and the binary orbit is appropriately inclined.

As mentioned above, in addition to wind absorption, the non-detection of radio pulsation in some HMGBs could also be due to the pulsar emission beam not pointing towards Earth. Following Emmering \& Chevalier (1989), the radio beaming fraction of pulsars can be written as follows:
\begin{eqnarray}
  f\left( \chi \right) &=&\left( 1-\cos \chi \right) +\left(\frac{\pi}{2}-\chi \right) \sin \chi,
\end{eqnarray}
where the radio beaming angle is (Kijak \& Gil 2003; Takata et al. 2011)
\begin{eqnarray}
  \chi &\simeq& 0.01265\nu _{\rm{GHz}}^{-0.13}P^{-0.35}\dot{P}_{-15}^{0.035}\ {\rm rad},
\end{eqnarray}
and $\dot{P}_{-15}=\dot{P}/10^{-15}$ is the period derivative. Obviously, for young pulsars with faster rotational speeds and/or larger period derivatives, the radio beaming fraction would be larger.
In substituting the parameters of PSR B1259-63 into the above equation with $P=0.0477625\ {\rm s}$ and $\dot{P}=2.27875\times10^{-15}\ {\rm s\cdot s^{-1}}$, we can estimate the radio beaming fraction with $f\simeq38\%$ at the frequency of $\nu=15\ {\rm GHz}$.
For LS 5039 and other HMGBs, if the compact objects are indeed energetic non-accreting pulsars similar to PSR B1259-63, we would expect more than one-third of them with the radio beaming pointing towards us.
However, if the unknown object in LS 5039 is a magnetar, as suggested by Yoneda et al. (2020), the radio beaming fraction would be much smaller, that is, $f\simeq10\%$ with $P\sim9\ {\rm s}$ and $\dot{P}\sim3\times10^{-10}\ {\rm s\cdot s^{-1}}$. To confirm the nature of the compact objects in LS 5039 and other HMGBs, further observations and deep analyses from radio to $\gamma$-ray data are needed.

\begin{acknowledgements}

We are very grateful to the referee for the careful reading of the text and many valuable suggestions and comments to improve the manuscript significantly.
This work is supported by the National SKA Program of China (grant No. 2020SKA0120300), the National Key Research and Development Program of China (grant No. 2020YFC2201400), the National Natural Science Foundation of China (grant Nos. 11822302, 11833003, U1838102), and the China Postdoctoral Science Foundation (grant No. 2020M682392).
\end{acknowledgements}

\end{document}